\documentclass{aa}
\usepackage{amsmath,amsfonts,amssymb}

\usepackage{graphicx}

\usepackage[british]{babel}

\usepackage{url}

\begin{document}

\title{Understanding local dwarf spheroidals and their scaling relations
under MOdified Newtonian Dynamics} 

\titlerunning{Dwarf spheroidals in MOND}

\author{X.  Hernandez$^{1,2}$ \and S. Mendoza$^1$ \and  T. Suarez$^1$ 
  \and  T. Bernal$^1$}

\offprints{X. Hernandez, S. Mendoza}

\institute{Instituto de Astronom\'{\i}a,
      Universidad Nacional Aut\'onoma de M\'exico, AP 70-264, 
      Ciudad Universitaria, Distrito Federal 04510, M\'exico
    \and
      GEPI, Observatoire de Paris, Meudon Cedex, France \\
  \email{xavier,sergio,tsuarez,tbernal@astroscu.unam.mx}
}

\date{\today}

\abstract
{}
{{We use a specific form of the interpolation function in the MOND formalism, which optimally
accounts for the internal structure of dwarf spheroidal (dSph) galaxies, to explore the consequences 
it has on the scaling relations seen in these systems.}}
{{The particular form of the interpolation function we used leads to a law of
gravity that does not degrade the good fit
of the MOND proposal on galactic scales, and in fact, slightly
improves the accordance with observations at dSph scales. This formalism
yields a good description of gravitational phenomena
without the need of invoking any still undetected and
hypothetically dominant dark matter, in the weak field regime probed
by local dSph galaxies.}} 
{{Isothermal equilibrium density
profiles then yield projected surface density profiles for the local dSph
galaxies in very good agreement with observational determinations, for
values of the relevant parameters as inferred from recent observations
of these Galactic satellites.  The observed scaling relations for these
systems are also naturally accounted for within the proposed scheme,
including a previously unrecognised correlation of the inferred mass-to-light
ratios of local dSph's with the ages of their stellar populations, which is natural
in modified gravity schemes in the absence of dark matter.}}
{ The results 
shed some light on the form that the MOND interpolating function may have 
in the most challenging regime, which occurs at moderate accelerations and intermediate
mass--weighted lengths.}


\maketitle

\section{Introduction} \label{intro}

If one wants to explore options where the existence
of dark matter is not a necessity, one must consider variations in the
law of gravity on galactic and extra-galactic scales where dark
matter has been proposed to account for the observed dynamics. The best-studied
such proposal is the MOdified Newtonian Dynamics (MOND) hypothesis of Milgrom (1983), which has
been shown to account for the rotation curves of spiral galaxies
(e.g. Sanders \& McGaugh 2002), naturally incorporating the Tully-Fisher
relation, see for example Milgrom (2008a) and references therein.

Abundant recent publications of the velocity dispersion measurements
of stars in the local dwarf spheroidal (dSph) galaxies, the extended and flat rotation
curves of spiral galaxies, the high dispersion velocities of galaxies in
clusters, the gravitational lensing due to massive clusters of galaxies,
and even the cosmologically inferred matter content for the universe,
are all successfully modelled using MOND, not as indirect evidence of a dominant 
dark matter component, but as direct evidence of
the failure of the current Newtonian and general relativistic theories of
gravity, in the large-scale or low-acceleration regimes relevant to the
above. Some recent examples of the above are Milgrom \& Sanders (2003),
Sanders \& Noordermeer (2007), Nipoti et al. (2007), Famaey et al. (2007),
Gentile et al. 2007), Tiret et al. (2007) and Sanchez-Salcedo (2008).

The range of galactic dynamical problems treated with the MOND formalism
has extended from the first-order gravitational effects of rotation curves and velocity 
dispersion measurements to cover a wide range of more subtle problems.
With their role in limiting the sizes of satellites and establishing in MOND escape
velocities for satellites or galaxies subject to an external acceleration field, tidal
forces were studied by Sanchez-Salcedo \& Hernandez (2007). Wu et al. (2008) calculated the escape
velocity for the Milky Way using both MOND and dark matter, and
conclude that the LMC appears as
bound from both points of view, in spite of the recently determined high proper motion for
this object, with slightly better fits using the MOND prescription.
Sanchez-Salcedo et al. (2008) looked at the thickness
of the extended HI disk of the Milky Way from both angles and
find a somewhat better fit to observations in the MOND theory. 
Sanchez-Salcedo et. al (2006) and Nipoti et al. (2008) examined
the problem of dynamical friction in dSph galaxies, comparatively assuming
Newtonian gravity or MOND, with decay timescales for globular clusters
being somewhat shorter in MOND. This is a potential problem, unless one allows
large initial orbital radii for the observed globular clusters beyond the current extent of the
stellar populations.

Going to cosmological scales, Skordis et al. (2006) have studied the
cosmic microwave background, Halle et al. (2008) looked at the problem of
the cosmic growth of structure, Zhao et al. (2006) studied gravitational
lensing of galaxies, and Angus et al. (2007) and Milgrom \& Sanders (2008)
studied dynamics of clusters of galaxies. All of the above find the 
MOND description of the problem to be a viable option
within the observational errors of the relevant determinations,
despite the need to include some unseen mass on galaxy cluster scales. 
Options for the above include the cluster baryonic dark matter proposed by Milgrom (2008b), or the inclusion of
neutrinos, whether active as studied by Sanders (2003) and Angus et al. (2007)
or sterile as proposed by Angus et al. (2010), always within plausible expectations.
Also, in spiral galaxies, although within observational errors, rotation curves have been reported 
by Gentile et al. (2004) and Corbelli \& Salucci (2007) to be
somewhat less well-fitted assuming the MOND prescription than when using cored dark matter haloes.

Observationally, it appears that MOND fares as well as dark matter in
accounting for measured dynamics, in all but the smallest scales,
the case of local dSph galaxies remains the most controversial.
Published studies using MOND, which typically assume the deep MOND regime,
sometimes find a different value for the acceleration scale of the
theory for different dSph galaxies, when fitting detailed
dynamical models to the data (Lokas 2001), or that $M/L$ ratios remain
higher than those of stellar populations, making dark matter necessary
even under the MOND formulation, e.g., Sanchez-Salcedo \& Hernandez (2007). Still, this
last point remains somewhat controversial. Angus (2008) suggests that the
MOND proposal will hold for local dSph galaxies, once the effects of
tidal disruption are adequately incorporated.

Regarding corresponding theoretical developments, numerous alternative
theories of gravity have recently appeared (Bekenstein 2004, Sanders
2005, Sobouti 2007, Bruneton \& Esposito-Farese 2007, Zhao 2007, Arbey
2008), now mostly grounded on geometrical extensions of general relativity
and field theory, which lead to laws of gravity in the Newtonian limit
that in the large-scale or low-acceleration regime reduce to the 
MOND prescription fitting formula.

The MOND theory is characterised by (i) a low acceleration regime where dynamics mimic
the presence of dark matter, (ii) a high acceleration regime where Newtonian
gravity is recovered, and (iii) a somewhat ill-defined transition region. Variants
of MOND are defined by the form of the interpolation function, $\mu(x)$, which
mediates the transition between the two limit regimes, see Eq.(1).

We show how the MOND prescription with a suitable interpolation function, such as
the one used by Bekenstein (2004) and used by Famaey \& Binney (2005), can
be written as the addition of both low acceleration limit MOND and
Newtonian contributions to the acceleration on all scales. The large
differences in scales and magnitudes of the two acceleration terms ensure
that the addition does not spoil the good match with observations on galactic
scales. We are trying to look for a suitable MOND interpolation function 
for the local dSph galaxies, but not necessarily one that works on all scales.

This, as it is the regime of the local dSph galaxies, the one which has proven
most difficult for MOND, and the one where the high-quality
recent observations now available allow for a relatively clean test. We
find that this interpolation function not only yields acceptable $M/L$
values for all dSph galaxies without the need for any dark matter in all cases,
but also provides a natural explanation for all the scalings seen in these Galactic satellites.  
An interesting correlation between the \( M / L \) ratios inferred and the ages of
the stellar populations in local dSph galaxies is found,
and naturally accounted for in modified gravity schemes where stars
alone account for the gravitational force. 

However, on solar system scales, Sanders (2006) has shown this particular
interpolation function to be incompatible with limits on the variations of Kepler's
constant. Thus, the good accordance we find with observations for dSph's
must be understood as evidence in favour of the particular interpolation function we test, at 
the weak field limit relevant for these systems. What a full interpolation function for
MOND should be is a more complicated issue. One, however, which has to be addressed bearing in
mind the results presented here.

Section (2) gives a brief summary of the interpolation function first
explored by Bekenstein (2004) and gives equilibrium isothermal
configurations.  These are then used in section (3) to model the local
dSph galaxies and obtain $M/L$ ratios, which are found to be consistent
with those typical of stellar populations, at the ages of the different
dSph systems. Section (4) includes an exploration of the most conspicuous
scaling relations for local dSph galaxies, finding that
Bekenstein's interpolation function naturally accounts for all 
features found. Our conclusions are summarised in section (5).

\section{A particular form of the interpolating function in the 
   MOND prescription}

In terms of the acceleration $\boldsymbol{g}$ felt by a test particle, the MOND 
proposal is (see e.g. Milgrom 2002)

\begin{equation}
  \mu \left(\frac{g}{a_{0}} \right) \boldsymbol{g} = \boldsymbol{g}_{N}
\label{the-equation}
\end{equation}

\noindent where $ \boldsymbol{g}_{N}$ is the acceleration assigned by
Newtonian gravity, $a_{0}$ an acceleration scale of the theory, and the
interpolating function $\mu(x:=g/a_0)$ is an unspecified function that
reduces to unity for large values of its argument, and to its argument,
for small values of it. In this way, Newtonian gravity is recovered for
large values of $g$, and in the deep MOND regime one obtains

\begin{equation}
g=\left( a_{0} g_{N} \right)^{1/2}.
\end{equation}

Strictly speaking, MOND is a theory defined only at the
limit values for the acceleration.  The value of the constant $a_{0}$
has been reasonably determined by calibrating the deep MOND regime
through observations of rotation velocity curves of large spiral galaxies
(e.g. Sanders \& McGaugh 2002).  However, the details of the  transition
region between the two regimes (i.e. the function $\mu$)
have proven harder to establish reliably. This is in part natural, as
it is not a scalar parameter, but a functional dependence that one is
looking for here. A number of variants for this function have been proposed (e.g. Famaey \&
Binney 2005, Zhao 2007, Sanders \& Noordermeer 2007), but apparently
suitable dynamical systems in the transition region are hard to come
by. Astrophysically, to first approximation one tends to find either
systems where no dark matter is needed (stellar and planetary systems,
globular clusters, vertical dynamics of disk galaxies, galactic bulges, and
elliptical galaxies), or systems where dark matter is massively dominant
(dSph Galactic satellites, rotation curves of spiral galaxies, galaxy
groups and clusters, and cosmological observations). Also, the details
of any empirically inferred transition region, $\mu(x)$, are sensitive to the
details of the baryonic system one is looking at, such as the assumed
mass-to-light ratios, gas dynamics, or orbital anisotropy of stars,
most of which play a marginal role when in the deep MOND regime.
As pointed out already, this ill-defined transition is cumbersome to
handle in attempts to reproduce the MOND phenomenology through simple
GR extensions e.g. Sobouti (2007) or Capozziello et al. (2007).

  Bekenstein (2004) shows that his relativistic extension of T$e$V$e$S in
the appropriate non-relativistic limits yields the interpolating function

\begin{equation}
  \mu(x) = \frac{ \sqrt{ 1 + 4x } - 1 }{ \sqrt{ 1 + 4x} + 1 }.
\label{interpol}
\end{equation}

\noindent This particular interpolating function converges to the right
limits as \( x \longrightarrow 0, \ \infty \) and has a very peculiar
property.  Direct substitution of equation~\eqref{interpol} into
the absolute value of relation \eqref{the-equation} yields

\begin{equation}
  g=g_{N} + \left( a_{0} g_{N} \right)^{1/2} .
\label{eq0}
\end{equation}

\noindent  This equation can be thought of as a generalised gravity recipe
described by the addition of two terms, the first a standard Newtonian
acceleration term and the second the MOND limit acceleration term.  Seen in this
way, equation~\eqref{eq0} changes its acceleration behaviour limiting
cases to a scale limiting behaviour.  Indeed, for the case of a test
particle on a gravitational field produced by a central mass \( M
\), located at a distance distance \( R \) from it,
equation~\eqref{eq0} can be written as

\begin{equation}
  g= -\frac{ G M }{ R^{2} } -\frac{ \left( G a_0 \right)^{1/2}  \, 
    M^{1/2} }{ R },
\label{eq1}
\end{equation} 

\noindent where \( G \) is Newton's constant of gravity.  Seen in this way,
equation~\eqref{eq1} converges to Newtonian gravity for sufficiently small
\( R / M^{1/2} \)'s and reproduces the MOND strong limit for sufficiently 
large \( R / M^{1/2} \)'s,
and so the acceleration limits are now ``scale--weighted'' limits.  It is
worth noting that equation~\eqref{eq1} is certainly the simplest
modification to Newtonian gravity once the weak acceleration MOND regime is known.

  Let us now rewrite equation~\eqref{eq0} in such a way that the
scale weighting  becomes clearer. To do so, recall
that $g_{N}=-\nabla \phi_{N}$, where \( \phi_N \) represents the standard 
Newtonian potential and $g=-\nabla \phi$, where \( \phi \) is the scalar
potential of the gravitational field. With this, equation~\eqref{eq0} can be
written as

\begin{equation}
  g=g_{N} \left(1+ \chi \right),
\label{eq3}
\end{equation}

\noindent where $\chi := (a_0 / G )^{1/2} \, (R/M^{1/2})$. In what follows
we use the standard values of $a_{0}=1\times 10^{-8} \, \mathrm{cm}\,
\mathrm{s}^{-2}$ (Milgrom 2008a)  and $G = 4.5 \times 10^{-39}
M_{\odot}^{-1} \, \mathrm{s}^{-2} \, \mathrm{kpc}^{3}$, in units suitable for galactic
applications.

As already mentioned, we are interested in the form of $\mu(x)$ at scales of the local
dSph galaxies. That the particular $\mu(x)$ we are testing is incompatible with solar 
system dynamics implies that a complete $\mu(x)$ MOND function is probably complex, 
that is, if one wants to treat the MOND formalism as more than just an empirical 
phenomenological description of gravitational physics on galactic scales.
On the other hand, some
researchers (see e.g. Bekenstein 2006 and references therein) claim
that the Pioneer anomaly can be explained by the MOND hypothesis.  This assumption
can prove wrong if uneven thermal radiation in the spacecrafts is found
(Toth \& Turyshev 2009), which could possibly come from the flyby anomaly
(Turyshev 2009). In other words, the MOND formalism does not appear
to be relevant on solar system scales, where General Relativity and
Newtonian Gravity have proven correct (e.g. Turyshev \& Toth 2009,
Anderson et al. 2002). Still, as discussed by Milgrom (2009), it is possible that
current tests on solar system scales cannot reach definitive
conclusions on the MOND interpolating function.

For globular clusters, with  $M=10^{5-6} M_{\odot}$, $R=2-10
\times 10^{-3} \, \mathrm{kpc}$, we get values for $\chi$ of between $0.1$ and
$0.01$, and the correction becomes smaller than the errors and
uncertainties in the observational determinations for the values of radii
and masses for globular clusters. For elliptical galaxies and bulges,
with masses going from about $10^{9}$ to $10^{11} M_{\odot}$ and radii
of between 0.5 and 10 kpc, $\chi$ is about 0.1. We thus see that
the correction to gravitational dynamics due to the proposed inclusion
of a second term in equation~\eqref{eq3} is small enough to have remained
undetected in galactic systems where no gravitational anomaly is found and
where dynamics are consistent with Newtonian gravity, in the absence of
any dark matter.

On the other hand, for systems where the presence of dark matter is inferred, given that this is
always required to be dominant, adding the Newtonian term onto the MOND proposal generally provides 
a negligible
contribution. For example, for the Galactic disk at the solar radius,  
$M=5 \times 10^{10} M_{\odot}$ and $R=8.5 \, \textrm{kpc}$ give $\chi=2$, consistent with an inference of 
about $50\%$ dark matter within the solar circle. In going to the outskirts of the Milk Way, we go to
$R=100 \textrm{kpc}$, hence $\chi=20$. The system is either totally dominated by dark matter, or it is in a regime
where the second term in equation ~\eqref{eq3} almost fully determines the dynamics. We see that the distinct 
power-law dependences ensure that the Newtonian term completely dominates at small values of $\chi$, while the 
opposite holds for large values of $\chi$, with a necessarily narrow transition region.


From the form of equation~\eqref{eq1}, it would be tempting to
add the following term of the \( \sim 1 / R \) series, a further
(perhaps positive) constant term, which would result in an $\chi^{2}$
additive term in equation~\eqref{eq3}. If chosen suitably small, for
the same reasons as given above, it would have no measurable effects on all
but the largest scales, perhaps as a tool to model the cosmological
constant.   This extension certainly does not correspond to MOND, since
no interpolation function \( \mu(x) \) can be constructed.  However,
in such terms, equation~\eqref{eq3} might be interpreted as a series
expansion of a more fundamental underlying gravity law, only the first
terms of which we have begun to appreciate empirically as observations
probe increasingly higher $\chi$ regimes, typically corresponding to
increasingly larger scales. The introduction of the constant term in
equation~\eqref{eq1} and its calibration from cosmology, however, must
be done within a generalised GR framework.

\subsection{Equilibrium configurations}

To test at the regime which has resulted most troublesome
for the MOND formalism, we go to the now very well studied dSph galaxies of the local group.
To compare against local dSph galaxies, we require the derivation of
equilibrium configurations for a population of self-gravitating stars.
We begin by noting that the validity of Newton's theorems for spherically
symmetric matter distributions, that the contribution to the force felt
by an observer due to external shells vanishes, and that the contribution
of all shells interior to the observer is equivalent to concentrating all
mass interior to the observer at the centre, also holds under the present
proposal. The two theorems depend on the numerator of the
Newtonian term in equation~\eqref{eq1} for a fixed solid angle fraction of a thin
shell scaling with the second power of distance to the shell, as the
denominator does. For the second term in equation~\eqref{eq1}, the numerator scales
with $\sqrt(R^{2})=R$, as does the denominator, assuring the validity of
Newton's theorems for spherical mass distributions (e.g. see Bekenstein \&
Milgrom 1984).

We now write the equation of hydrostatic equilibrium for a polytropic equation of state $p=K \rho^{\gamma}$:

\begin{equation}
K \gamma \rho^{\gamma-2} \frac{ \mathrm{d} \rho}{ \mathrm{d} r} = - \nabla
  \phi = - \frac{GM(r)}{r^{2}} - \frac{ \left( a_0 / G \right)^{1/2}
  M(r)^{1/2} }{r}.
\end{equation}

\noindent Since  $\rho = (4 \pi r^{2})^{-1} \mathrm{d}M(r)/\mathrm{d}r$,
where \( M(r) \) is the mass of the configuration at a radius \( r \),
and going to isothermal conditions, $K=\sigma^{2}$ with $\gamma=1$, the
previous equation can be written as

\begin{equation}
\sigma^{2}\left[ \left( \frac{ \mathrm{d}M}{\mathrm{d}r} \right)^{-1}
  \frac{d^{2}M}{dr^{2}} -\frac{2}{r} \right] = -\frac{G M(r)}{r^{2}}
  -\frac{ \left( a_0 / G \right)^{1/2} M(r)^{1/2}}{r}.
\end{equation}

In the above second-order differential equation for $M(r)$,
the constant isotropic velocity dispersion for
the population of stars has been written as $\sigma$. We have
taken advantage of the correspondence between hydrostatic equilibrium
polytropic configurations and self gravitating stellar systems, which
does not depend on the particular form of the Poisson equation, only
on the validity of the isothermal condition for the stellar population
(see e.g. Binney \& Tremaine 2008). Taking initial conditions $M(r)\rightarrow 0$ and $ \mathrm{d}M/
\mathrm{d}r = 4 \pi r^{2} \rho_{0}$ for $r \rightarrow 0$, a constant
central density $\rho_{0}$, we can now solve equation (8) through a
numerical finite differences scheme, subject only to two input conditions,
a value for $\sigma$ and a value for $\rho_{0}$.

Notice that neglecting the first term on the right hand side of equation (8),
the limit behaviour for $M(r)$ for large values of $r$ when the weak limit 
of the MOND formalism dominates, is characterised by a finite total mass.
In contrast to what happens in Newtonian systems
where isothermal configurations have an infinite total mass, 
isothermal self-gravitating configurations will
be naturally bound in mass, under the proposed gravity law.  
As a result, there will also be a well-defined
and finite half-mass radius, $R_\textrm{hm}$, to characterise the resulting
equilibrium configurations. Also, as is the case using the MOND prescription, the total
mass of the configuration is expected to scale with $\sigma^{4}$.
From the dimensional mass scale $<M> \sim \sigma^{4} \left( a_0 
/ G \right)^{-1}$, one
is led to expect an analogue to the Tully-Fisher relation for spheroidal
galactic systems, down to the dSph regime.

Figure~(1) gives a plot of $\log( \rho / M_{\odot} \, \textrm{pc}^{-3})$
vs $\log(R/\textrm{kpc})$ for a sample numerical solution to equation
(8) for $\sigma=7 \, \textrm{km} \, \textrm{s}^{-1}$ and $\rho_{0}=0.1 \,
M_{\odot} \, \textrm{pc}^{-3}$. We obtain a well-defined resulting
total mass, $M_{tot}$, of $3\times 10^{6} M_{\odot}$, and a final volume
half-mass radius of $0.39 \, \textrm{kpc}$.  We see the expected decay
of the density profile at large radii, slightly faster than $r^{-3}$. It
is this type of density profiles that is used in the next section to 
model the local dSph galaxies.

\begin{figure}
\begin{center}
\includegraphics[height=6.51cm,width=8.01cm]{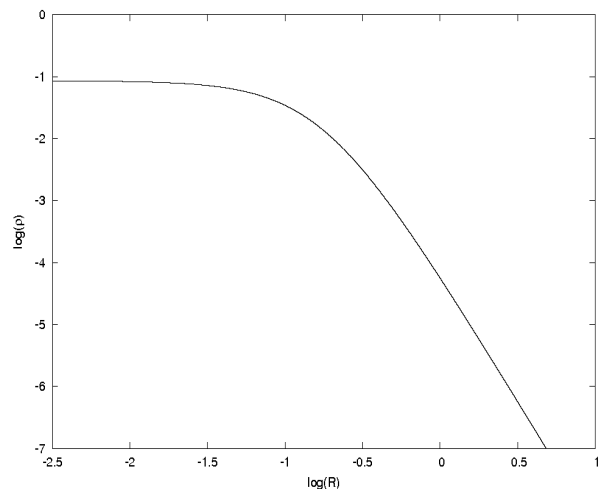}
\end{center}
  \caption{A sample isothermal equilibrium density profile for
$\rho_{0}=0.1 M_{\odot} pc^{-3}$ and $\sigma=7 km/s$, resulting in a
total mass of $3\times 10^{6} M_{\odot}$ and a volume half-mass radius of
$0.39 \textrm{kpc}$. The density is measured in units of $M_{\odot}
pc^{-3}$ and the distance in units of $\textrm{kpc}$.}
\end{figure}

\section{Local dwarf spheroidal galaxies}

As discussed above, once values for the velocity dispersion and the
central density are given, numerically solving equation (8) yields
the full equilibrium density profile. To now model dSph galaxies, we
take values for the reported velocity dispersions for these systems,
which thus fix one of the two parameters of the model. A resulting
density profile is then integrated along one dimension to yield a
surface density profile, from which the projected half-mass radius is
measured. This provides a second constraint; we then vary the value
of the input central density to ensure that the resulting half-mass
radius $R_{\rm hm}$ matches the reported half-light radius
$R_{\rm{hl}}$ of a given dSph.  Observational determinations of the
velocity dispersion and the half-light radius of a certain dSph galaxy
then fully determine the model. Dividing the total mass for the
resulting model $M_ {\rm tot}$ by the reported total luminosity
$L_{\rm tot}$ of a dSph, yields a mass to light ratio for the model. 
Values of the half-light radii and total luminosities were taken from
Gilmore et al. (2007), who summarise published results from the
references given in Table (1), except for Ursa Minor, where we took the
revised value for $L_ {\rm tot}$ from Palma et al. (2003).

\begin{center}
\begin{table*}
\flushleft
\begin{minipage}{130mm}
  \caption{Basic properties and resulting $M/L$ ratios for the sample of dSph galaxies.}
  \begin{tabular}{@{}llllllll@{}}
  \hline
  \hline
   Galaxy & $\,\,\,\sigma\,[\rm{km/s}]\,\,\,\,\,\,\,\,$ & $R_{\rm hl}\,[\rm{kpc}]\,\,\,\,\, $ & $L_{\rm tot}\,\times10^{5}L_{\odot}$
  & $(M/L)_{A}\,\,\,\,\,\,\,\,\,\,\,$ & $(M/L)\,\,\,\,\,\,\,\,\,\,\,\,\,$ & $<\chi>\,\,\,\,\,\,\,\,\,\,$  &  
Age of Youngest \\
      & & & & & & &  Component [Gyr]\\
 \hline
 & & & & & & \\
 Carina     & $7\pm\,1.8$   & 0.290 & 4.3   & $5.6^{+5.2}_{-2.9}$  &  $6.8^{+8.3}_{-4.6}$   & 7  &  3   \\
 & & & & & & \\
 Draco      & $8\pm\,1.5$   & 0.230 & 2.6   & $43.9^{+29}_{-19.3}$  &  $17.0^{+13.9}_{-8.9}$ & 4.1  & 10   \\
 & & & & & & \\
 LeoI       & $8\pm\,1.2$   & 0.330 & 48.0  & $0.7^{0.65}_{-0.3}$   &  $1.0^{+0.6}_{-0.44}$  & 5.7  & 2     \\
 & & & & & & \\
 Sextans    & $7\pm\,1.0$   & 0.630 & 5.0   & $9.2^{+5.3}_{-3.0}$   &  $6.3^{+4.2}_{-2.8}$   & 13.4 & (2-6) \\
  & & & & & & \\
 Fornax     & $12\pm\,1.3$  & 0.400 & 150.0 & $1.4^{+0.45}_{-0.35}$&  $1.4^{+0.6}_{-0.4}$   & $>3.4$   & (2-3)  \\
  & & & & & & \\
 Sculptor   & $9.5\pm\,1.7$ & 0.160 & 22.0  & $3.7^{+2.2}_{-1.4}$  &  $3.4^{+2.1}_{-0.7}$   & $>2.2$ & $>5$     \\
  & & & & & & \\
 LeoII      & $6\pm\,1.4$   & 0.185 & 7.0   & $1.85^{+2}_{-1.1}$   &  $2.3^{+2.4}_{-04}$    & 5.5  & 6.5    \\
  & & & & & & \\
 Ursa Minor & $8\pm\,2$     & 0.300 & 5.8   & $5.8^{+6.5}_{-3.6}$   &  $8.0^{+9.8}_{-5.1}$   & $>5.3$ & 12     \\
\hline

\end{tabular}

\end{minipage}

\begin{flushleft} $(M/L)_{A}$ gives the values for the mass to light
ratios calculated by Angus (2008), and $M/L$ those
in this study under the proposed model. Total luminosities (in the $V$
band) and half-light radii are from Wilkinson et al. (2006), Wilkinson et
al. (2004),
Koch (2007a), Kleyna et al. (2004), Walker et al (2006), Mateo (1998),
Coleman et al. (2007) and Irwin \& Hatzidimitriou (1995), for the galaxies in the Table, in
the order given, as summarised in Gilmore et al. (2007). Velocity dispersions
are from adjusting a constant value to the data of Angus (2008). $<\chi>$
gives the average value of the parameter $\chi$ as defined in section (2),
which measures the relative relevance of the Newtonian and 
the low acceleration limit of the MOND formalism terms in equation (6). The final column gives an estimate of the age of
the youngest stellar population present in each of the systems,
for Carina, Ursa Minor, LeoI, and LeoII from Hernandez et al. (2000),
for Fornax from Coleman \& de~Jong (2008), for Draco from Aparicio et al. (2001), for
Sextans from Lee et al. (2003), and for Sculptor from Babusiaux et al. (2005).

\end{flushleft}

\end{table*}

\end{center}

  Although other stellar systems show velocity dispersion profiles that
generally decay as the radial coordinate increases, e.g. globular
clusters, the case for observed dSph galaxies is different, with
these systems showing essentially flat velocity dispersion profiles,
e.g. Gilmore et al. (2007). In cases where some drop in the velocity
dispersion profile is observed, this typically occurs towards the edge of
the galaxy, affecting only a very small percentage of the total mass of
the system. The constancy of these observed velocity dispersion profiles
validates the use of equation (8) under isothermal conditions for 
modelling local dSph's. We have
not at this point attempted a more detailed modelling considering
non-isothermal conditions, for example including orbital anisotropy,
although such models are now common in models of dSph galaxies under
Newtonian (e.g. Lokas 2002) or MOND frameworks (e.g. Angus 2008)
and could be included in subsequent analysis. We point to the recent work by
Gilmore (2007), in which isothermal conditions are used to model
local dSph galaxies assuming Newtonian gravity.

  Values of $\sigma$ were taken from adjusting a constant level
to the $\sigma$ profiles of Angus (2008), which are flat to a very
good approximation, with the exception of Ursa Minor that shows
a significant decrease with increasing radial distances. This galaxy
also shows internal structure in phase space at small radii, which might
partly account for the steep increase in $\sigma$ seen by Angus (2008)
towards the centre. We have taken a value for $\sigma$ representative
of the situation at around the half-light radius.  If a strong
radial dependence of $\sigma$ in Ursa Minor were confirmed, however, this system
would have to be excluded from the present sample, as it would then 
conflict with the simple isothermal modelling we are performing here.

Table (1) gives our results for the 8 best-studied local dSph galaxies.
The values we obtain for $M/L$ are mostly comparable to those obtained by
Angus (2008) for the same systems, but a slight systematic decrease
is evident, consistent with having included a further force term, the
Newtonian component, which reduced inferred $M/L$ ratios even further.
In general, the values of the parameter $<\chi>$ we found, calculated from
the resulting model total mass as $(a_0/G)^{1/2}(2^{1/2}R_{\rm hl}/M_{\rm
tot}^{1/2})$, explain why the addition of the Newtonian term is required
to fully account for the dynamics. Further, the wide range of values of
$<\chi>$ shown in the table explains why a different $a_{0}$
has sometimes been found for various dSph's, when fitting dynamics using
only the weak acceleration limit of the MOND formalism, e.g. Lokas (2001).

  The case where the effect of the second term in equation~\eqref{eq3} is
greatest is that of Draco, where the compact configuration and relatively
high $\sigma$ values allow the Newtonian term to figure somewhat more,
leading to a strong decrease in the inferred $M/L$ of between 8.1 and
30.9, compared to the Angus (2008) results of $M/L$ between 22.6
and 72.9. With the possible exception of Fornax and Sculptor,
cases where the reported King core radius serves only as a lower limit
for the projected $R_{\rm hl}$, the value for the parameter $<\chi>$
for Draco is the lowest in the sample. This naturally explains why it
is here that our results for $M/L$ differ most and show a significant
reduction when contrasting with the results of Angus (2008), who takes a \( \mu\) 
function that rapidly converges to the low acceleration limit of the MOND prescription, as 
this is the parameter that determines the relative importance of the Newtonian
and low acceleration limit MOND terms in equation~\eqref{eq3}.

  The $1\sigma$ confidence intervals of our results are only lower
bounds, as their calculation includes only uncertainties in the adopted
values of $\sigma$ and not observational errors in total luminosities
and half-mass radii. Also, freedom in the light profile functional
fitting would increase our confidence intervals slightly, e.g., the
reported $R_{\rm hl}$ values for Ursa Minor, Fornax and Sculptor are only
lower bounds, because the quantity available for those galaxies is the core
radius from King profile fits. Still, this last is only a second-order
effect, since our inferred $M/L$ values are not very sensitive to the values
of $R_{\rm hl}$ used, provided this does not change by a large factor
(see the following section).

  As expected, the main determinant of the resulting total mass of a
numerical solution of equation (8), is the input $\sigma$. We find, as
expected from the situation of the low acceleration limit of the MOND prescription, a strong
correlation, with the resulting scaling for all galaxies being well
described by

\begin{equation}
M_{\rm tot} = \frac{ ( 3.36 \sigma^2 )^2 } { a_0 G } .
\end{equation}

\noindent This could be seen as an extension of the Tully-Fisher relation down
to the smallest galactic scales, a natural consequence of the model being
explored.

  Considering the large $M/L$ values natural for old stellar populations,
e.g., Queloz et al. (1995) and Romanowsky (2003), who find $M/L$ ratios for old
stellar populations as high as 7 or 8, our new results are now at the
limit of compatibility with $M/L$ ratios for old naked stellar populations
for all the well-studied local dSph galaxies. This holds comfortably for
all galaxies except Draco, where compatibility is found only just at the
$1\sigma$ level for our lower limits on the confidence intervals.
Our model thus yields equilibrium isothermal configurations able to
explain the observed dynamics of the local dSph galaxies, without the
need for dark matter.

\begin{figure}
\begin{center}
\includegraphics[height=6.51cm,width=8.01cm]{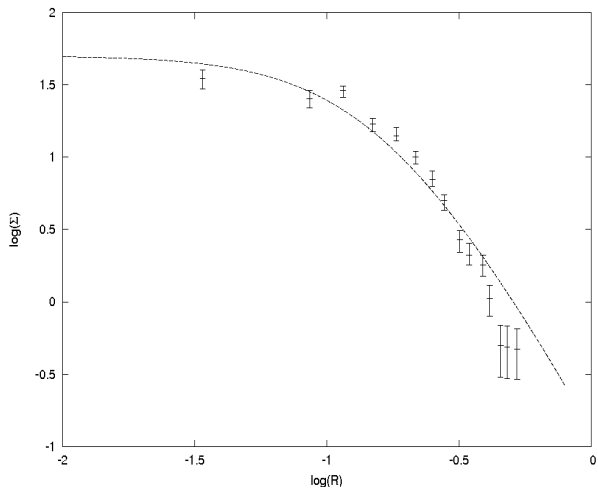}
\end{center}
  \caption{Comparison of our projected surface density profile for an
equilibrium isothermal solution to equation (8), having input $\sigma$
and projected $R_{\rm hm}$ as the observationally determined values for
LeoII, dashed curve. For comparison, we also give the star counts surface
density profile for Leo II of Coleman (2007). Both profiles have been
normalised to the same total luminosity.}
\end{figure}

  A consistency check of the above interpretation is available from the
comparison of the $M/L$ ratios we found and the age of the youngest
stellar population found in each of the galaxies, as inferred from the
direct studies of the observed CMDs of the galaxies in question, given
in the last column of Table (1). It is reassuring that the highest values
for our inferred $M/L$ ratios, for Ursa Minor and Draco coincide with the
galaxies showing no star formation over the past 10 Gyr, the oldest
galaxies in the sample, while the lowest values for $M/L$ are obtained
for Leo I and Fornax, the youngest galaxies in the sample with stellar
populations as young as 2 Gyr. For the remainder of the sample, galaxies
showing their youngest stars at intermediate ages, we find
the intermediate values for $M/L$. The interpretation of the dynamics
under the proposed model thus accords with the natural increase
in the $M/L$ ratios of stellar populations due to the ageing of stars
and the consequent build up of black holes, neutron stars, and white dwarfs.
This correspondence is natural in any modified gravity scheme where the stars 
alone are responsible for the dynamics, but has to be thought of as a fortuitous
coincidence under the dark matter hypothesis.

  As a second consistency check, we now show the resulting projected
surface density mass profile for a model of the Leo II galaxy, normalised
by the total luminosity of that system. We construct the surface
brightness profile using our model for an equilibrium isothermal system
having velocity dispersion and projected $R_{\rm hm}$ (assumed equal to
the observed $R_{\rm hl}$), equal to those observed for Leo II. This is
given in Figure~(2), where we have also plotted the observation for the
actual surface density light profile of Leo II, from the star counts
analysis of Coleman et al. (2007), out to the radius where measurements fall
below the background noise level, both normalised to the same total
luminosity. We note that the error bars of Coleman et al. (2007) are only a
lower estimate of the confidence intervals for this comparison, as they
only refer to the errors in the measured star counts and not to full
surface density profile inferences. A very good agreement is evident,
showing the proposed models are a good, fully self-consistent
representation of the dynamics and both the integral and spatially
resolved light distribution in the well-studied local dSph galaxies.

\begin{figure}
\begin{center}
\includegraphics[height=6.51cm,width=8.01cm]{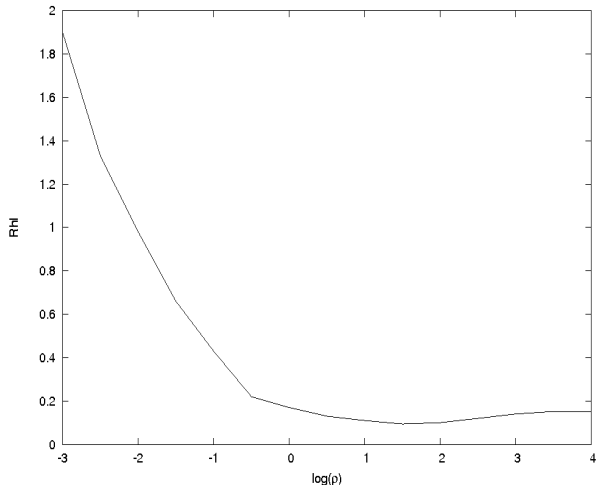}
\end{center}
  \caption{Dependence of the resulting projected half-mass radius in kpc,
against the assumed central density in $M_{\odot} \rm{pc}^{-3}$, for
isothermal equilibrium configurations under the proposed MOND formalism,
at fixed $\sigma=10\rm{km/s}$. A very broad region where $R_{\rm hm}$
remains almost constant is evident, at precisely the level found to define
the minimum $R_{\rm hl}$ values for observed dSph galaxies, which have
typical values of $\sigma \simeq 10 \rm{km/s}$. }
\end{figure}

\section{Scaling relations}

\begin{figure}
\begin{center}
\includegraphics[height=6.51cm,width=8.01cm]{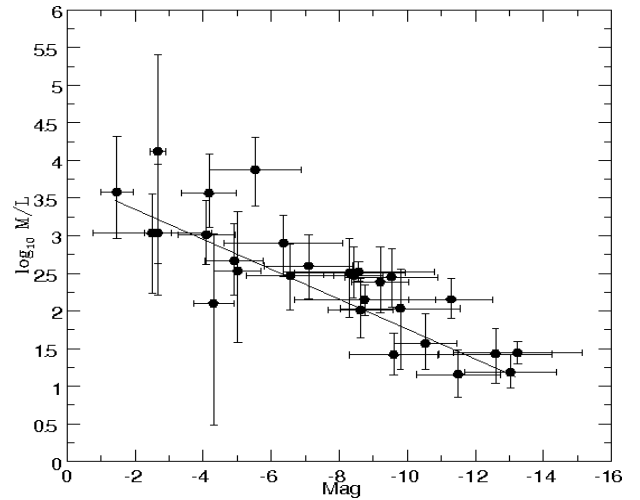}
\end{center}
  \caption{Logarithms of the Newtonian $M/L$ values of local dSph galaxies
against total observed $V$ band magnitudes, from data as summarised in Walker et al. (2009a).
The straight line is not a fit, but gives the Newtonian $M/L$ values that would be assigned to
isothermal equilibrium populations of stars having a constant intrinsic
stellar $M/L$ ratio of 5 and a constant half-mass radius of $300\rm{kpc}$,
consistent with what is observed, and to what the proposed model yields
for equilibrium configurations having velocity dispersions of $10\rm{km/s}$,
as the observed dSph show, according to the model presented here.
Galaxies in ascending order of luminosity, with original references: 1- Sgr(23, 24), 2- Fornax (1,2),  
3- And II (17, 18), 4- Leo I (1, 5), 5- Cetus (17, 22), 6- Sculptor (1, 2), 7- And XV (20, 21), 8- Leo II (1, 6), 
9- Tucana (25, 26), 10- Sextans (1, 2), 11- Draco (3, 4), 12- Carina (1, 2), 13- Cven I (3, 10), 14- Umi (1), 
15- And IX (19), 16- Leo T (3, 10, 14), 17- Hercules (3, 11), 18- Bootes 1 (3, 8), 19- UMa I (3, 8), 20- Leo IV (3, 10), 
21- Cven II (3, 10), 22- Leo V (12, 13), 23- UMa II (3, 10), 24- Coma (3, 10), 25- Willman 1 (3, 8), 
26- Bootes 2 (3, 9), 27- Segue 2 (16), 28- Segue 1 (3, 15).
References: 1- Irwin \& Hatzidimitriou (1995), 2- Walker et al. (2009c), 3- Martin et al. (2008), 
4- Walker et al. (2007), 5- Mateo et al. (2008), 6- Koch et al. (2007b), 7- Martin et al. (2007), 
8- Koch et al. (2009), 9- Simon \& Geha (2007), 10- Aden et al. (2009), 11- Belokurov et al. (2008), 
12- Walker et al. (2009b), 13- Irwin et al. (2007), 14- Geha et al. (2009), 15- Belokurov et al. (2009),
16- McConnachie \& Irwin (2006), 17- Cote et al. (1999), 18- Chapman et al. (2005), 19- Ibata et al. (2007),
20- Letarte et al. (2009),  21- Lewis et al. (2007), 22- Ibata \& Irwin (1997), 23- Majewski et al. (2003),
24- Saviane et al. (1996), 25- Fraternali et al. (2009).}
\end{figure}

We now turn to the scalings shown by the dSph galaxies in our sample.
Firstly, we show in Figure~(3) the behaviour of the equilibrium isothermal
configurations we are solving for, in terms of the resulting projected
$R_{\rm hm}$ as a function of the input value of $\rho_{0}$, at a constant
value of $\sigma = 10 \rm{km/s}$. The remarkable feature of Figure~(3) is
that, after decreasing as $\rho_{0}$ increases, when $\rho_{0}$ reaches a
value of about $0.3 M_{\odot} \rm{pc}^{-3}$, the resulting $R_{\rm hm}$
stops changing and settles at relatively constant value of around
$150\rm{pc}$, below which it does not fall further. This is qualitatively
reproduced at all values of $\sigma$, with only small changes in the
minimum values of $R_{\rm hm}$, for $\sigma$ in the range of values
observed for local dSph galaxies. This is interesting, as it offers a
natural explanation for local dSph galaxies showing a
minimum lower value for their projected half-light radii of around
$150\rm{pc}$, with most lying around a factor of 2 above this critical
limit, as noticed by Gilmore et al. (2007) and seen from Table (1). Indeed,
all our models constrained by the observational values of $\sigma$ and
$R_{\rm hl}$ for the local dSphs, occur within the flat region of 
the $R_{\rm hm} \rm{vs.} \rho_{0}$ space.

  We can now try to understand the assigned Newtonian values of $M/L$ for
local dSph galaxies and the scalings they show with total luminosity or
absolute magnitude, e.g., Mateo (1998). The assigned
Newtonian values of $M/L$ will never be far from (e.g. Gilmore et al. 2007)

\begin{equation}
(M/L)_{N}=\left( \frac{10 \sigma^{2} R_{\rm hl}}{G}\right) \left( \frac{1}{L_{\rm tot}} \right) .
\end{equation}

  Given our results in the previous section, or alternatively taking the 
observed $R_{\rm hl}$ of around $300\rm{pc}$ as an empirical fact, we can
evaluate the above equation to first order at $R_{\rm hl}=0.3$. Also, for
an average old stellar population, we can take a constant $M/L$ value of
5 as a representative value for the local dSphs, as our inferences yield,
and replace $\sigma^{2}$ in equation (9) for the corresponding value
through equation (10) to yield

\begin{equation}
(M/L)_{N}=\left( \frac{3 \,\sqrt{5}}{3.36} \right) {\left( \frac{a_0}{G} \right)} ^ {1/2} \left( \frac{1}{L_{\rm tot}} \right)^{1/2} .
\end{equation}

\noindent Introducing the absolute magnitude $M_{V} =-2.5 \rm{log}(L_{\rm tot}) + 4.83$
and taking the logarithm of the above equation gives

\begin{equation}
\rm{log}(M/L)_{N}=3.77 + 0.2 M_{V}.
\end{equation}

  Figure~(4) now gives a plot of equation (12), superimposed on recent
determinations of the Newtonian $M/L$ values for a larger sample of
local dSph galaxies, using reported values of L to calculate the values plotted on the
x-axis, and of L, $\sigma$, and $R_{hl}$ to calculate Newtonian $M/L$ values through equation (10). Error
bars give extreme $1\sigma$ confidence intervals, including uncertainties in all the
parameters used. Data as summarised in Walker et al. (2009a) (using the revised version of the
data for the classical dSphs, as given in Walker et. al 2009d),
and references therein. The good match is evident, extending for over 13 orders
of magnitude in $M_{V}$, which is even more remarkable because the trend found extends
towards the smallest newly discovered dSph systems.

  We see that the expectations of the model for the inferred Newtonian
$M/L$ values of local dSph galaxies, which are modelled as equilibrium isothermal
solutions to equation (8) and having typical radii of close to
$0.3 \rm{kpc}$, agree very well with independent measurements. The
small scatter in Figure~(4) beyond observational error bars is compatible with the variations in the
actual intrinsic $M/L$ values for the individual galaxies, and variations
of $R_{\rm hl}$ around the average $0.3\rm{kpc}$ used in equation (12),
which is not a fit to the data, but a first-order estimate within our
prescription. A very simple explanation for the $(M/L)_{N}$ values
assigned to all well-studied dSphs is thus naturally afforded by our
assumptions for the most average intrinsic $M/L$ and $R_{\rm hl}$ values,
without the need of invoking complex astrophysical processes (e.g. tidal
disruption, strong outflows, etc.) or fine-tuning any parameters. 
To first order, one can understand the well-established scalings of figure (4) 
as the extension of the Tully-Fisher relation, which appears in the MOND formalism,
down to the dSph regime.

\section{Conclusions}\label{ccl}

We used a MOND interpolation function that can be seen as the
addition of both the Newtonian acceleration and the low acceleration limit of the
MOND formalism on all
scales (Bekenstein 2004).  For the local dSph galaxies,
where the application of the MOND formalism has been most controversial, we show that 
isothermal equilibrium configurations characterised by well-defined
finite total masses and half-mass radii result, giving $M/L$
values in agreement with naked stellar populations. As a result,
no dark matter is now needed.  The observed scalings
in $R_{\rm hl}$ and assigned $(M/L)_{N}$ values as a function of total
magnitudes are explained naturally, including the previously unrecognised correlation
of  resulting $M/L$ ratios with the relative youth of the stellar populations of the
individual dSph galaxies, for the sample studied. This last is a natural consequence of
using the MOND prescription, but an odd coincidence under the dark matter hypothesis.

Fixing the optimal $\mu(x)$ for the MOND prescription in the dSph regime through the arguments presented
here constitutes an interesting condition on this function in the regime studied, and it
provides a constraint on any global $\mu(x)$ MOND function. 

\begin{acknowledgements}
We would like to thank Benoit Famaey, Moti Milgrom, and Jacob Bekenstein
for fruitful comments in connection to a previous version of this
article.  We thank an anonymous referee for comments that were 
useful towards reaching a more balanced and complete presentation. 
This work was supported in part through two DGAPA-UNAM grants
(PAPIIT IN-113007-3 and IN-114107).  SM and TB gratefully acknowledge
support from DGAPA (IN119203-3) at the Universidad Nacional Aut\'onoma
de M\'exico (UNAM).  SM acknowledges financial support by
CONACyT~(26344).  TB acknowledges support from CONACyT (207529).
\end{acknowledgements}

\bibliographystyle{aa}
\bibliography{dwarfRLE}

\end{document}